\documentstyle[11pt,oldlfont]{article}
\renewcommand{\theequation}{\thesection.\arabic{equation}}
\def\Re{\rlap{\rm I}\mkern3mu{\rm R}}
\def\Zed{\rlap{\rm \mbox{\sf Z}}\mkern3mu{\rm \mbox{\sf Z} }}
\def\a{\alpha}

\def\d{\delta}
\def\e{\epsilon}

\def\f{\varphi}

\def\lb{\label}

\def\m{\mu}
\def\n{\nu}
\def\c{\cite}

\def\r{\rho}

\def\esp{\mbox{\Large{e}}}
\def\Fb{\mbox{\boldmath$F$}}
\def\Wb{\mbox{\boldmath$W$}}

\def\st{\mbox{\boldmath$*$}}

\def\S{\mbox{\boldmath$\Sigma$}}

\def\Ab{\mbox{\boldmath$A$}}
\def\xib{\mbox{\boldmath$\xi$}}
\def\ub{\mbox{\boldmath$\ub$}}

\def\ub{\mbox{\boldmath$u$}}
\def\etab{\mbox{\boldmath$\eta$}}
\def\p{\partial}

\def\th{\theta}

\def\we{\mbox{\scriptsize \boldmath$\wedge$}}
\def\wed{\dot{\we}}
\def\L{\mbox{\boldmath$L$}}

\def\U{\mbox{\boldmath$U$}}
\def\Bb{\mbox{\boldmath$B$}}
\def\Gb{\mbox{\boldmath$G$}}

\def\il{\mbox{\boldmath$i$}}

\def\li{\mbox{$\mathcal{L}$}}
\def\ex{\mbox{\boldmath$d$}}
\def\cex{\mbox{\boldmath$D$}}

\def\bi{\bibitem}

\def\B{\begin{equation}}
\def\E{\end{equation}}

\begin{document}

\thispagestyle{empty}

\begin{flushright}   hep-th/9809109\\

                    LPTENS 98/36 \end{flushright}

\vspace*{0.5cm}

\begin{center}{\LARGE { On Superpotentials and Charge Algebras of Gauge Theories}}

\vskip1cm

S. Silva

\vskip0.2cm

Laboratoire de Physique Th{\'e}orique CNRS-ENS\\ 
24 rue Lhomond, F-75231 Paris Cedex 05, France\footnote{Unit\' e propre du CNRS, associ\' ee \` a l'Ecole Normale Sup\' erieure et \` a l'Universit\' e de Paris-Sud.}
\vskip1.0cm

\begin{minipage}{12cm}\footnotesize

{\bf ABSTRACT}
\bigskip 

We propose a new ``Hamiltonian inspired" covariant formula to define
(without harmful ambiguities) 
the superpotential and the physical charges associated to a gauge
symmetry. The criterion requires the variation of the
Noether current not to contain 
any derivative  terms in $\partial_{\mu }\delta \f$.

The examples of Yang-Mills (in its first order formulation) and 3-dimensional Chern-Simons theories are
revisited and the corresponding charge algebras (with their central extensions in the Chern-Simons case) are computed in a straightforward way.

We then generalize the previous results to {\it any}
$(2n+1)$-dimensional non-abelian Chern-Simons theory for a particular
choice of boundary conditions. We compute explicitly the
superpotential associated to the non-abelian gauge symmetry which is
nothing but the Chern-Simons Lagrangian in $(2n-1)$ dimensions. 
The corresponding charge algebra is also computed.
However, no associated central charge is found for $n \geq 2$. 

Finally, we treat the abelian p-form Chern-Simons theory in a similar way.
 
\bigskip
\end{minipage}
\end{center}
\newpage

\section{Introduction}
\setcounter{equation}{0}

For a long time, gauge symmetries have been present in all the areas of theoretical physics. The simple question of what are all the associated Noether conserved charges due to these local symmetries has been treated 
partially in a lot of papers since the first work of E. Noether \c{No}
(see also \c{Ju} and \c{JS} and references therein). The striking
point is that these charges, unlike the global-symmetry Noether
charge,  generically can (and must) be computed as an integral over the boundary of a spacelike hypersurface (say at spatial infinity). 
Thus what we need, instead of the usual $(D-1)$-form Noether current, is a $(D-2)$-form, also computed locally as a functional of the fields at spatial infinity. 
These $(D-2)$-forms are the superpotentials, whose systematic derivation was carried out for the first time in the gravitational framework by Bergmann's school \c{Be} in the 50's, and for the general p-forms theories by B. Julia \c{Ju} in 1981. 
As an example, let us recall the simplest superpotential, the one associated to the $U(1)$ gauge symmetry of Maxwell theory, namely, the field tensor $F^{\m\n}=\p^\m A^\n -\p^\n A^\m$.

All the above results have been reviewed and developped in \c{JS} in great detail (see also \c{BCJ} and \c{BFF}) using the ``cascade equations", which are nothing but a compact way to write down the computations of \c{Be} and \c{Ju} plus the Noether identities \c{No}. 
This will be the starting point of the present article that is why the techniques of \c{JS} will be recalled in the preliminaries, section 2.

As was also pointed out in \c{JS}, the cascade equations formalism suffers from an ambiguity in the choice of a surface term which can only be cured by a case by case prescription. 
In most examples, this prescription is so natural that the cascade
equations technique gives an easy way to compute the
superpotential. The Chern-Simons theories are exceptions and some
``natural" choices can give incorrect superpotentials, as we shall show
in section 4\footnote{The same problem happens in supergravities, see
for instance \c{HJS}.}.

This ambiguity issue, plus the fact that the first computed superpotentials were not in general gauge invariant (e.g., Freud's superpotential for the gravitational field) contributed to their unpopularity\footnote{Although this last problem was solved for the gravitational case in the beginning of the 60's by Rosen and Cornish \c{RC}, and more recently in a completely general way by Katz, Bi\u{c}{\'a}k and Lynden-Bell \c{KBL} (in the later work, one should restrict the discussion to a neighbourhood of $\infty$).}.
Other related problems, as the famous factor of 2 between the mass and the angular momentum of Kerr black hole computed with the Komar superpotential\footnote{Again, this problem is treated in great detail in \c{JS} (see also \c{Ka}, \c{Ch} and \c{KBL}). 
The conclusion was that the use of Komar's superpotential is simply inconsistent with the asymptotically flat boundary conditions of the Kerr black hole. The correct superpotential \c{Ka} can also be found with the cascade equations \c{JS}.}, 
motivated people to look for another techniques to compute the conserved charges associated to gauge symmetries. At the cost of loosing explicit covariance (Hamiltonian formalism), Arnowitt, Deser and Misner studied the problem of charges in the gravitational Einstein theory and found the celebrated ADM mass formula \c{ADM}, which is nothing but an asymptotic version of Freud superpotential. 
Analogously time and space are split in the Regge and Teitelboim \c{RT} unambiguous computation (up to a constant) of the asymptotic charges (superpotential) within the Dirac Hamiltonian formalism (see also the work of Brown and Henneaux \c{BH1}).

The question of how to reconcile such an unambiguous charge formula
(up to a constant) and the covariant Noether program for gauge
theories remained open. In section 3, we propose a simple formula for
computing the superpotentials which combines these two techniques,
when the equations of motion depend at most on first derivative of the
fields\footnote{This last supposition is not too restrictive at all
since it covers Yang-Mills and (super)-gravities in their first order
formulations, Chern-Simons theories, etc\dots}. It is nothing else than the natural covariant generalization of the Regge-Teitelboim program.
In fact, what will be proposed is a formula for the variation of the superpotential which does not depend on possible addition of a total derivative to the Lagrangian as in the cascade equations formalism. 
Generically, the integration of this equation can be carried out only
after imposing some boundary condition on the fields. As in the
Hamiltonian formalism, the integrability of the superpotential remains
an open problem.
Moreover, in
some cases, the gauge symmetry has to be fixed, see for instance the
diffeomorphism symmetry of Chern-Simons theories. Finally, to justify
our proposal, we study a {\it consistency test}; a general
proof that this {\it consistency} holds for all Lagrangians which are of the form $L=f^{\mu}
(\f) \partial_{\mu} \f + g (\f)$ is presented in the second part of the
appendix.

The proposed formula will be used in section 4 to treat the Yang-Mills
(in its first order formulation) and
Chern-Simons-type theories\footnote{The case of gravity and
supergravity will be treated in \c{JS2} and \c{HJS} respectivelly. We
can just anticipate that the superpotentials derived in that way are
in complete agreement with the standard Hamiltonian results of Regge
and Teitelboim.}. 
In the first two examples we recover old results and illustrate how
the new superpotential formula works. The next two examples, namely
the higher dimensional non-abelian and the abelian p-form Chern-Simons
theories, are treated in detail for a particular choice of boundary
conditions. 
The superpotentials corresponding to the (non)-abelian gauge symmetry
are computed and found to be simply proportional to the Lagrangian
densities of the respective theories in two dimensions less. The
algebra is also computed and it is shown that no central charge appears
for $n \geq 2$ (with our {\it particular} choice of boundary conditions), where $D=p+n
(p+1)$, $p$ and $D$ being the form degree of the basic field and the
spacetime dimension respectivelly. Finally, The superpotentials associated to the diffeomorphism symmetry are also discussed.

We are not going to give again the motivations for studying these Chern-Simons topological field theories. We will just mention here a few related topics which will be mentioned in due time such as 2+1-(super)gravities, BTZ black hole and its microcanonical entropy computation, generalized higher dimensional (super)gravities, topological field theories, anomaly calculations, AdS/CFT Maldacena conjecture etc...

\bigskip

\section{Preliminaries}
\setcounter{equation}{0}

Let $L(\f,\p \f) $ be a Lagrangian which for simplicity just depends on  the fields $\f$ and their first
 derivatives. Let $\xi^a (x)$ ($^a$ denotes the Lie-algebra gauge
index) be the local parameter of a gauge symmetry $\d_\xi$ under which
the action $\int L(\f,\p \f)$ is invariant, that is $\d_\xi L =\p_\m S_\xi^{\m}$.  In that case the Noether construction tells us that:

\B \d_\xi \f\  \frac{\d L}{\d \f} =  \p_\m J_\xi^\m\lb{offcas}\E

Where 

\B J_\xi^\m := S_\xi^{\m}-\d_\xi \f\  \frac{\p L}{\p \p_\m \f}  \lb{nocu} \E

\noindent is the parameter-dependent Noether current and $\frac{\d L}{\d \f}$ are the Euler-Lagrange equations of $\f$ and an abstract summation over all the fields of the theory is understood.

The variation of the field $\d_\xi \f$ can be written as:

\B \d_\xi \f = \xi^a \Delta_a(\f) + \p_\n \xi^a \Delta^\n_a(\f)  \lb{tran1} \E

This is just the generic case where the transformation of the field depends at most on one derivative of $\xi^a (x)$. A more general case can be treated in the same way.

As is well known, the surface term $S_\xi^{\m}$ of (\ref{nocu}) (and so $J_\xi^\m$) is not uniquely determined by the Noether construction. In fact the addition of an exact term, say $\p_\n S_\xi^{[\m\n]}$, is undetectable\footnote{ This problem exists for any Noether current (even for global symmetries) as soon as one of the fields of the theory, generically the gauge field, does not vanish at the boundary.} 
since the only object that we can compute from the Lagrangian is
$\p_\m S_\xi^{\m}$.  What we need then is to make a choice of such a surface term, say for example

\B S_\xi^\m = \xi^a \Sigma_a  ^\m (\f) +  \p_\n \xi^a \Sigma_a  ^{\m\n} (\f) \lb{tran2} \E

It is very important to keep in mind that this choice is {\it a
priori} arbitrary. We will come back to this point at the end of this
section but for the moment let assume that $S_\xi^\m$ is given by
(\ref{tran2}) and continue with the (off-shell) abelian cascade
equations. Following  \c{Ju} and \c{JS}, we use (\ref{tran1}) and
(\ref{tran2}) in (\ref{offcas}) to obtain that:

\B \left( \xi^a \Delta_a + \p_\n \xi^a \Delta^\n_a  \right)  \frac{\d L}{\d \f} = \p_\m \left( \xi^a J^\m_a  + \p_\n \xi^a U^{\m\n}_a  \right) \lb{idno} \E

 where 

\B J_a ^\m := \Sigma ^\m_a - \Delta _a  \frac{\p L}{\p\p_\m\f}   \lb{defj} \E

\B U^{\m\n}_a  := \Sigma^{\m\n}_a  - \Delta^\n _a \frac{\p L}{\p\p_\m\f} \lb{cur2} \E

Now, let us use the abelian restriction
\B \xi^a (x) := \e (x)  \xi_0^a(x)  \label{abelres}\E
in (\ref{idno}). $\e (x)$ is an arbitrary function (with no gauge group index dependence) and $\xi_0^a(x)$ is fixed but spacetime dependent. What we obtain is then 

\B \e\ \d_{\xi_0} \f\frac{\d L}{\d \f} + \p_\n \e\ \xi_0^a \Delta^\n_a\frac{\d L}{\d \f}   = \p_\m \left( \e\ J_{\xi_0}^\m  + \p_\n \e\  U^{\m\n}_{\xi_0}  \right) \lb{idnoab} \E

Where $J_{\xi_0}^\m$ is given by (\ref{nocu}) and 

\B U^{\m\n}_{\xi_0} := \xi_0^a\ U^{\m\n}_a  \lb{nosu} \E

The abelian (off-shell)-cascade equations are obtained by making use of the arbitrariness of $\e (x)$ and its derivatives. Thus, from (\ref{idnoab}) we derive:

\B \e(x) \hspace{.3in} \left[ \d_{\xi_0} \f\ \frac{\d L}{\d \f} =  \p_\m J_{\xi_0}^\m \right] \lb{idcas1} \E

\B \p_\m \e(x) \hspace{.3in} \left[ \xi_0^a \Delta^\m_a\ \frac{\d L}{\d \f} = J_{\xi_0}^\m + \p_\n U^{\n\m}_{\xi_0}  \right] \lb{idcas2} \E

\B \p_\m \p_\n \e(x)  \hspace{.3in} \left[ U^{\m\n}_{\xi_0} = 0 \right]  \Rightarrow  
U^{\m\n}_{\xi_0} = U^{[\m\n]}_{\xi_0}\lb{idcas3} \E

These three cascade equations encode all the Noether information of local symmetries, that is, the on-shell conservation of Noether current (equation (\ref{idcas1})), the existence on-shell of an antisymmetric superpotential for the Noether current 
(equation (\ref{idcas2}) and (\ref{idcas3})) and the Noether
identities for the equations of motion (noting the fact that the
derivative of the lhs of equation (\ref{idcas2}) is equal to the lhs of (\ref{idcas1}) by virtue of the antisymmetry of superpotential, equation (\ref{idcas3})).

Now equation (\ref{idcas2}) can be rewritten (by making use of the
antisymmetry property of the superpotential $U^{\m\n}_{\xi_0}$, and
omitting the $_0$ subscript) as:

\B J_\xi^\m = \xi^a \Delta^\m_a \frac{\d L}{\d \f} +
\p_\n U^{\m\n}_\xi \lb{startpoint} \E

That is, the Noether current can always be written on-shell (the first
term in the rhs then vanishes) as the divergence of a superpotential defined locally in terms of the fields by equations (\ref{nosu}) and (\ref{cur2}). 
This superpotential is naturally defined modulo an exact term which does not contribute since the physical charge (see equation (\ref{loccha}) below) is always computed on a closed manifold. Equation (\ref{startpoint}) will be our starting point in section 3.

Moreover, the physical conserved charge is given by the integral of the superpotential at spatial infinity\footnote{The question of what is the precise decrease will not be treated here.} \c{JS}:

\B Q(\xi) = \int_{B_{\infty}} U_\xi^{\m\n} dS_ {\m\n}\lb{loccha} \E
This follows from the physical requirement that the current
$J^{\mu}_{\xi}$ has to vanish at infinity:

\B \int_{\Sigma_{\infty}} j^{\mu}_{\xi} dS_{\mu} = 0 \label{vanj} \E
\noindent where $\Sigma_{\infty}$ is a $(D-1)$ timelike hypersurface
located at infinity.

Of course the $\xi$'s are not arbitrary (so the number of conserved charges is not {\it a priori} infinite) but are dictated by the boundary conditions \footnote{ Actually the case of non viscous fluid mechanics whithout boundary is quite different. 
The $\xi$'s are determined in that case by the fields or thermodynamic functions of the theory in the bulk. The striking point is that there is an infinite number of known independent $\xi$'s in any even space dimension (the enstrophies) and only one in odd space dimensions (the helicity), for details, see \c{JS} and references therein.} . 
For example, if Dirichlet boundary conditions ($\d \f = 0$) are used, the allowed $\xi$'s would be given by the solutions of the asymptotic equation:

\B \lim_{r \rightarrow \infty} \d_\xi \bar{\f} = 0 \lb{bcres} \E

Where $\bar{\f}$ is our particular solution to the field equations.

\bigskip

Now, it is time to come back to the arbitrariness of the surface term $S^\m$ (\ref{tran2}). We need to make a choice for this term: for example $\xi^\m L$ is the usual choice for the diffeomorphism invariance of gravity theory, and $0$ for the internal gauge invariance of Yang-Mills theory. 
What was also emphasized in \c{JS} is that once such a choice is made, the addition of a total derivative to the Lagrangian changes the associated superpotential. In other words, if the boundary condition changes (say Dirichlet or Neumann), the associated Lagrangian and superpotential change and so does the physical charge. 
There the example of gravity was treated (in the new Affine gauge formulation) and it was explicitly shown that the superpotential associated to metric-Dirichlet type boundary condition (i.e. $\d g_{\m\n} = 0$) was the Katz superpotential, which is nothing but a covariant version (with some background correction added, see \c{KBL} and references therein) of ADM mass formula and reduces to Freud or Nester-Witten ones.
The choice $\d g_{\m\n} = 0$ selects Einstein $\Gamma\ \Gamma$ action against Hilbert's covariant action.

The drawback of this method is that we need some intuition to give the right surface term (\ref{tran2}). In the case of general relativity or Yang-Mills theory it is more or less natural and as we said, the associated superpotential gives the expected charges.

In the case of a gauge theory where all the charges are defined at the boundary, this kind of ambiguity due to the choice of surface term is as important as the superpotential itself. 
In fact, a straightforward calculation shows that if we add to $S_\xi^\m$ the exact term $\p_\n \left(S_a^{[\m\n]} \ \xi^a\right)$, where $S_a^{[\m\n]}$ is any functional of the fields, the superpotential $U_\xi^{\m\n}$ of (\ref{nosu}) is shifted precisely by $S_a^{[\m\n]} \ \xi^a$. A particular choice can cancel the whole superpotential. 

This concludes our review of the needed results of \c{JS} and gives the motivations for the following section.

\section{Superpotential from equations of motion}
\setcounter{equation}{0}

The point now is that in the case of Chern-Simons theories the
``natural" choice for the surface term is not so obvious\footnote{The
same is true for supergravities, see \cite{HJS}.} as we will see in the following examples of section 4. 
We will then propose an ``Hamiltonian inspired" definition of the
superpotential which depends on the chosen boundary conditions but
does not require a guess for $S^{\mu }_{\xi}$ of (\ref{tran2}) but rather determines it. We will restrict our proposal to theories formulated in
a first order form. Concretely that means (in our definition) that
the {\it equations of motion depend at most on first derivative of the fields}. 

\bigskip

In the Hamiltonian formalism (see  Regge and Teitelboim \c{RT} and also \c{BH1}) the
 generators of proper gauge symmetries $G_a$ (namely the first class
constraints) have to be supplemented with a boundary term:

\B G_Q = \int_\Sigma \xi^a G_a + \int_{\p \Sigma} Q [\xi] \lb{rt1} \E

Where $\Sigma$ is some (D-1)-dimensional spacelike hypersurface with boundary
$\p \Sigma$. The (variation of the) charge $Q[\xi]$ is computed by
requiring (using the chosen boundary conditions) the functional
derivative of $G_Q$ to be coming from the bulk without boundary term

\B \d G_Q = \int_\Sigma \d \Phi \frac{\d \left( \xi^a G_a \right)}{ \d \Phi} \lb{rt2} \E
 
The charge or boundary term $Q[\xi]$ in (\ref{rt1}) precisely
compensates the boundary term arising from the difference between $\d
(\xi^a G_a)$ and the rhs of (\ref{rt2}). Here $\Phi$ stands for bulk
coordinates and momenta at fixed time.

\bigskip

Let us come back to the Lagrangian point of view. Despite the {\it a priori}
arbitrariness in the choice of surface
term (see the end of section 2), equation (\ref{startpoint}) tells us that whatever choice is
made, the difference between the Noether current and the
divergence of the superpotential will always be well defined (and will
vanish on-shell), namely the first term of the rhs of
(\ref{startpoint}). Moreover the integral of the superpotential has to be
the conserved charge. So, the structure of (\ref{startpoint})
namely $J_\xi^\m = \xi^a \Delta^\m_a \frac{\d L}{\d \f} + \p_\n
U^{\m\n}_\xi$ is analogous
to equation (\ref{rt1}). The Noether current corresponds to generator of the improper
symmetry $G_Q$, the equations of motion correspond to first class
constraints associated to the proper gauge symmetries $\xi^a G_a$ (which
vanish on-shell) and, in this analogy, the superpotentials correspond
to the boundary terms needed to make the variation of the generators
of improper symmetries well defined.

Then, let us differentiate equation (\ref{startpoint}) with respect to an arbitrary field variation $\d \f$, keeping the local parameter $\xi^a$ fixed:

\B \d J_\xi^\m = \d \f \ \frac{\d W_\xi^\m}{\d
\f}  + \p_\n V^{\m\n}_{\xi}(W_\xi^\m, \d \f) +\p_\n \d U^{\m\n}_\xi \lb{startpoint2} \E

Where

\B  W_\xi^\m:=  \xi^a \Delta^\m_a\ \frac{\d L}{\d \f}\lb{defw} \E

and $V^{\m\n}_{\xi}(W_\xi^\m,\d \f)$ is the usual surface term arising from the variation:

\B V^{\m\n}_{\xi}(W_\xi^\m,\d \f):= \d \f  \frac{\p W^\m_{\xi}}{\p \p_\n \f}  \lb{defv} \E

Remember that we supposed that the equations of motion depend only on
the first derivatives of the fields. We also made the extra supposition
that this remains true for $\Delta^\m_a$. 
The question of how to extend this to higher derivatives will not be
discussed here. The reason is that most of the physically interesting
theories can be reformulated in a first order form which satisfy the
above restriction by adding some auxiliary fields (see for instance
the example of Yang-Mills theories in section 4). Moreover, the
Hamiltonian formalism is a first order formalism (the fields and their
canonical conjuguates are independent). Thus it will be easier to
compare ou result with the Regge-Teitelboim procedure.

\bigskip 

{\it Our main proposal (ansatz) is to force} the variation of
the Noether current to be ``localizable'' in the sense that

\B \d J_\xi^\m = \d \f \ \frac{\d W_\xi^\m}{\d \f} \label{defj2}\E

\noindent in analogy with (\ref{rt2}). This condition gives an equation for the
variation of the superpotential (remember (\ref{startpoint2})) , namely:

\B \d U^{\m\n}_\xi = - V^{\m\n}_{\xi}(W_\xi^\m,\d \f) \lb{defu} \E

The superpotential is then obtained after (functional)-integration of
equation (\ref{defu}) at the spatial boundary (say for instance infinity
$B_\infty$), taking into account now the chosen boundary conditions. 

The antisymmetry in $\mu$ and $\nu$ of $V^{\m\n}_{\xi}$ and of the superpotential
defined respectively by (\ref{defv}) and (\ref{defu})  is explicitely shown in the first part of the appendix.

In this appendix we also give a {\it consistency test} to justify the proposal
(\ref{defu}). This {\it test} has to be satisfied for
all examples where (\ref{defu}) is used to compute
superpotentials. Moreover, we also give a proof for (\ref{defu}) in
the case where the Lagrangian is of the form  $L=f^{\mu } (\f)
\partial_{\mu } \f + g (\f)$.

\bigskip

Let us summarize the important points of this definition:

- As we emphasized, it is nothing but a covariant version of the
Hamiltonian procedure. In fact, instead of the generators of the
proper gauge symmetry (which are a subset of the equations of motion and so vanish on shell), we use all the equations of motion (see (\ref{defw})).

- This definition does not require the knowledge of the surface term added to the Lagrangian since it depends only on the field equations through $ W^\m_\xi$ (\ref{defw}). This can be helpful for computations in some cases (see for instance the higher dimensional Chern-Simons theories treated below) where the Lagrangian is quite complicated whereas the equations of motion are simple. 

- The sometimes difficult step is to integrate this equation by making use of the boundary conditions. In some cases, some gauge fixing procedure will be necessary, breaking down the explicit Lorentz covariance (as in the case of the diffeomorphism invariance of Chern-Simons theories).
In fact, it is not guaranted that equation (\ref{defu}) is at all
integrable, however it appears to be the case for all generic examples (see \c{JS2} for the gravity case). Moreover, equation (\ref{defu}) gives the expected superpotentials (charges) for these cases.

- This procedure defines a superpotential up to a ``constant'', which can
be adjusted so that the total charge vanishes for vacuum. By ``constant'' we mean an expression which is
fixed by the boundary conditions, ie any topological charge for instance.

- The antisymmetry of $U^{\m\n}_{\xi}$ in $\m$ and $\n$ (defined at
spatial infinity) follows from definitions (\ref{defv}) and
(\ref{defu}). This is proved in the appendix, see equation
(\ref{antisym}). 
  
- Equation (\ref{defu}) determines the variation of the superpotential. This can be used to compute the algebra of the corresponding charges. Let us take a particular gauge symmetry variation, say $\d_\eta \f$ and integrate (\ref{defu}) on the asymptotic boundary $B_\infty$ to obtain the resulting charge variation (see (\ref{loccha})), say $\d_\eta Q(\xi)$. 
In all the cases treated below, this quantity will be antisymmetric in $\eta$ and $\xi$. This will allow us to define a bracket between two charges $[Q(\xi),Q(\eta)] := \d_\eta Q(\xi)$, which will be homomorphic (up to non-trivial extensions) to the gauge commutator in the examples treated below. 
The Jacobi identity will also be automatically satisfied in the above
examples. Since we do not have at our disposal any theorem (analogous to the
nice one in the Hamiltonian formalism given some time ago by Brown and Henneaux \c{BH1}, see also \c{Ar} for the many point particles case) which guarantees a Lie algebra structure in a general way, we will simply conjecture it. In the following sections, we will give examples to illustrate our procedure.

\bigskip

\section{Some Examples}
\setcounter{equation}{0}

All the examples here will be treated in the language of differential forms. We will not repeat all the derivations of previous section since the correspondance is straightforward. For a good introduction to the variational techniques on differential forms, see \c{Th}.

Some of the superpotentials that will be derived here are well known from the litterature. The purpose is thus to illustrate formula (\ref{defu}) in specific examples and to compare with the results obtained with other methods.

\subsection{The Yang-Mills Theory}

Let $\L_{YM}:= Tr \left( \frac{1}{2} \Fb \we \st \Fb -\left( \ex \Ab +
\Ab \we \Ab\right)\we \st \Fb\right)$ be the Yang-Mills $(D)$-form
Lagrangian in a first order formulation. Here $\Fb:=\frac{1}{2}F_{\mu
\nu }^{a} \ex x^{\mu }\we \ex x^{\nu } T_{a}$ is as usual the
Lie-algebra valued curvature tensor but has to be treated as an
independent field (the $T_a$'s represent a basis of the Lie algebra and the Lie-indices are moved up and down by the Killing metric $K_{ab} = Tr \left( T_a T_b\right)$ ). The other field is the standard Lie algebra valued gauge potential $\Ab:= A_\m^a\ \ex x^\m\ T_a$. The theory is invariant under the local symmetry:
\begin{eqnarray}
\d_\xi \Ab &=& \cex \xi = \ex \xi + [ \Ab, \xi ]  \lb{ymtran}\\
\d_\xi \Fb &=& [ \Fb, \xi ] \label{ymtran2}
\end{eqnarray}

Where $\xi=\xi^a\ T_a$ is the local gauge parameter and $[\ ,\ ]$ the Lie-bracket.

The equations of motion for $\L_{YM}$ are given by:

\begin{eqnarray}
\frac{\d \L_{YM}}{\d \Ab} &=& -\cex \st \Fb \approx 0 \lb{ymeq}\\
\frac{\d \L_{YM}}{\d \Fb} &=& \st \Fb - \st \left( \ex \Ab +
\Ab \we \Ab\right) \approx 0 \label{ymeq2} 
\end{eqnarray}

Now what we need is the Hodge dual of $W^\m_\xi$ (see definition  (\ref{defw})), that is a ($D-1$)-form  which is in that case simply (see (\ref{defw}) and (\ref{tran1}), with (\ref{ymtran}) and (\ref{ymeq})):

\B \Wb_\xi = - \xi\ \cdot \cex \st \Fb \lb{dot} \E

Anytime a single dot appears in a formula (this can happen also on a
wedge product, e.g. $\Ab \wed \Fb$), take the totally symmetrized
(Lie)-trace of the corresponding expression (within the
parenthesis). This convention will simplify the heavy Lie-index
notation and will be used in the rest of the paper. 

Note that due to the fact that the gauge transformation law for the
curvature (\ref{ymtran2}) do not depends on the derivative of the
gauge parameter, its associated equations of motion (\ref{ymeq2}) do
not contribute to (\ref{dot}).

From the definition (\ref{defv}), we can compute 
 the Hodge-dual of $V^{\m\n}_{\xi} (W^\m_\xi, \d \f)$, and so the varied
superpotential following (\ref{defu}),  that is (remember
that $\Fb$ is still an independent field): 

\B \d \U_\xi =  \d \st \Fb\cdot\xi \lb{func} \E

To integrate such an equation we will use some Dirichlet boundary
condition on the connection, that is $\d \Ab=0$. In that case $\Fb$,
treated as an independent field, is not determined by the boundary conditions and is allowed
to freely fluctuate.
Note that the Yang-Mills theory in its first order formulation is in
the class of Lagrangian (\ref{Lag1or}) for which the {\it consistency test} given in
the appendix is automatically satisfied. In fact a straightforward
calculation shows that equation (\ref{corect}) (or (\ref{simcheck})) gives for the
Yang-Mills theory $\d_{\xi} \Ab \wed \st \Fb - \d \Ab \wed \d_{\xi}\st
\Fb $, which vanishes at infinity due to the above boundary conditions.

Now, since $\xi$ is fixed, equation (\ref{func}) can easily be integrated at
spatial infinity to give the well known result:

\B \U_\xi =  \st \Fb \cdot \xi \approx \st \left(\ex \Ab +\Ab \we \Ab
\right)\cdot \xi \lb{ymsup} \E

The last formula was obtained in \c{JS} with the cascade equations
technique in a second order formalism and with the obvious null choice
for the surface term $S_\xi^\m$ of (\ref{tran2}). Since the conserved
charge has to be evaluated on-shell, both result coincides. As
expected, the Hamiltonian formalism gives the same result.

The Lie algebra of the charges can be computed in a straightforward
way. Let us vary (\ref{ymsup}) with respect to a gauge transformation
parametrized by $\eta$, keeping $\xi$ fixed\footnote{The variation of $\xi$ under $\eta$ on the
rhs of the following equation (\ref{eqmanq}) appears as a formal
trick to compute the algebra, which will be useful in the following
examples. A direct calculation gives of course the same result:
$\d_\eta \left( \st \Fb_a \right) \xi^a = \left( f_{abc} \eta^b \st \Fb^c
 \right) \xi^a = \st \Fb^c [\xi,\eta]_c$.}:

\B \d_\eta  \U_\xi =  \d_\eta \left( \st \Fb \right) \cdot \xi = \d_\eta \left( \st \Fb \cdot \xi \right) -  \st \Fb \cdot \d_\eta ( \xi) \lb{eqmanq} \E

The first term of the rhs vanishes since it is a scalar of the
gauge group (this is not always the case, see for instance the
Chern-Simons case below where it generates a central extension). The
second term is just $-[\eta, \xi]$ times the curvature, that is, by (\ref{ymsup}):

 \B \d_\eta  \U_\xi = \U_{[\xi,\eta]} \E

As a final comment, we would like to repeat that the $\xi$'s or $\eta$'s are not arbitrary functions of the boundary coordinates but are restricted by condition (\ref{bcres}). The case where some matter current is added does not change the above results. Sometimes the $\xi$'s may be even more restricted, see \c{JS} and references therein.

\subsection{The Chern-Simons Theory}

Let us now take the Chern-Simons Lagrangian with its usual normalisation:

\B \L_{CS} = \frac{k}{4 \pi} Tr \left( \Ab \we \ex \Ab + \frac{2}{3} \Ab \we \Ab \we \Ab \right) \lb{cslag} \E

The associated equations of motion are then:

\B \frac{\d \L}{\d \Ab} = \frac{k}{2 \pi} \Fb  \lb{cseq} \E

\noindent Now $\Fb$ is not an independent field as in previous
example. It is given by the standard formula $\Fb =\ex \Ab +\Ab \we
\Ab $.

The interest in the Chern-Simons charge algebra started indirectly with the work of Brown and Henneaux \c{BH2} on 2+1 dimensional anti-De Sitter gravity, where they found that the asymptotic charge algebra was the conformal algebra in 2 dimensions equipped with a central extension. 
The equivalence of both theories (namely Chern-Simons and gravities in 2+1) found by Ach{\'u}carro and Townsend \c{AT} and Witten \c{W} allowed Ba{\~n}ados \c{B1} to recover this result by means of a twisted Sugawara construction based on the current (affine Kac-Moody) algebra (on $S^1$). 
This conformal central charge plays a key role in the determination by Strominger \c{St} of BTZ \c{BTZ} black hole entropy (see also the nice paper of Carlip \c{Ca} and references therein for related topics).

\bigskip

The 3-dimensional Chern-Simons theory is invariant under two local symmetries:

\bigskip

\underline{The non-abelian gauge symmetry}:

The transformation of the field is, as in the Yang-Mills case, given by (\ref{ymtran}). With the definitions (\ref{tran1}), (\ref{defw}) and (\ref{defv}) and by use of (\ref{cseq}) , equation (\ref{defu}) becomes:

\B \d \U_\xi = -\frac{k}{2 \pi} \d \Ab \cdot \xi \lb{arucs} \E

Remember here the ``dot convention" (section 4.1, equation (\ref{dot})) which simply means ``take the totally symmetrized Lie-trace".
One important point here is that we cannot fix all the $\Ab$'s at the boundary since they do not Poisson commute. That means that some $\Ab$'s will be arbitrary and so the rhs of (\ref{arucs}) does not vanish identically. 
We can however integrate this equation in a Lorentz covariant and local way\footnote{We again assume as in the rest of the paper that the gauge parameters do not depend on the fields of the theory.}:

\B \U_\xi = -\frac{k}{2 \pi} \Ab \cdot \xi \lb{ucs} \E

Unlike the Yang-Mills case, this expression is not gauge invariant. This will imply as we will see the existence of a central charge.

\bigskip

Here we would like to make some comments: 

- The coefficient in front of the superpotential (\ref{ucs}) depends
only on the normalisation of the Lagrangian. The corresponding
Hamiltonian result can be found in the work of Ba{\~n}ados \c{B1} using
Hamiltonian methods. What we point out here is that the canonical split (2+1 in this case) is not necessary, as the formula (\ref{defu}) shows.

- In a recent paper, Borowiec et al. \c{BFF} found this superpotential with some techniques very similar to the cascade equations formalism reviewed in section 2 (see also \c{Ju}, \c{JS} and \c{BCJ}). However as we alluded to in the introduction, the surface term choice suffers from an ambiguity which is not as natural to solve in that case as in the Yang-Mills or Gravity cases. 
In fact, the variation of (\ref{cslag}) under (\ref{ymtran}) may be the exterior derivative of $\frac{k}{4 \pi}   \xi \cdot \ex \Ab$, of $\frac{k}{4 \pi} \Ab \wed \ex \xi $ (this is the choice of \c{BFF}), or even the sum of the above plus $\a\ex (\Ab\cdot \xi)$ ($\a$ being an arbitrary constant). Of course, any of these choices changes the coefficient in front of the superpotential in an arbitrary way. 
The question is then to know what is the ``natural" choice for the surface term. Equation (\ref{defu}) gives an unambigous answer, namely equation (\ref{ucs}). This problem could be seen here just as a ``charge normalization" problem; however its consequences can be more drastic, as for the diffeomorphism symmetry.

- The Chern-Simons Lagrangian is of the type (\ref{Lag1or}). This
implies that the {\it consistency test} is automatically satisfied
(see appendix A.2), and thus, the use of (\ref{defu}) is fully
justified. In fact, equation (\ref{corect}) gives $\frac{k}{2\pi}
\delta \Ab \wed \delta_{\xi}\Ab$, which vanishes at infinity due to the
boundary conditions (\ref{cscond}-\ref{cscond2}).

\bigskip

Let us now vary (\ref{ucs}) with respect to a gauge transformation parametrized by $\eta$:

\B \d_\eta \U_\xi =-\frac{k}{2 \pi} \xi \cdot \d_\eta\Ab = -\frac{k}{2 \pi} \d_\eta  \left( \Ab \cdot \xi \right) + \frac{k}{2 \pi} \d_\eta(\xi) \cdot \Ab \lb{ctrick} \E

As in equation (\ref{eqmanq}) the second term on the rhs gives precisely $\U_{[\xi,\eta]}$. But contrary to equation (\ref{eqmanq}), the first term on the rhs is not a scalar of the gauge algebra due to the non homogeneous term in the transformation law of $\Ab$ (i.e. $\ex \eta$, see (\ref{ymtran})). This term is precisely the central charge of the algebra:

\B \d_\eta \U_\xi = \U_{[\xi,\eta]} - \frac{k}{2 \pi} \ex \eta \cdot \xi  \lb{alcs}\E

Let us recall finally how the boundary conditions can restrict the gauge parameter $\xi$.
For completeness, we give a quick overview of this important point (the generalization to higher dimensions will be done in the following section).

In general the variation of a Lagrangian gives rise to a boundary term which has to vanish for the corresponding choice(s) of allowed boundary conditions. For the Chern-Simons Lagrangian (\ref{cslag}), that means

\B \frac{k}{4 \pi} \int_{\Sigma_\infty}  \d \Ab \wed \Ab  = 0 \lb{cscond} \E

$\Sigma_\infty$ is a 2-dimensional asymptotic manifold which in this case is simply taken isomorphic to $\Re\times S^1$. 
We can then choose the usual cylindrical coordinates, that is $(t,r,\th)$, so that $\frac{\p}{\p t}$ is a time like vector, $\frac{\p}{\p r}$ is the normal vector of the infinite boundary $\Sigma_\infty$ and $\th$ parametrizes the asymptotic $S^1$ sphere. So  the above condition (\ref{cscond}) can simply be written as:

\B \d A_{[t} \cdot A_{\th]} = 0 \hspace{.3in} \mbox{at the boundary}
\label{cscond2}\E

We can then take the stronger condition $A_t=0$ as a possible boundary condition to satisfy the above equation\footnote{Other possibilities connected to gravity assumptions are $A_{\pm}= A_t \pm A_\th=0$ \c{BBO} or more generally  $ A_t-\omega (t) A_\th = 0$ \c{BG}.}. To be consistent with that choice, the allowed asymptotic gauge parameters have to satisfy (\ref{bcres}):

\B D_t \xi=\p_t \xi =0 \hspace{.3in} \mbox{at the boundary} \lb{cscon} \E

That is, the Lie algebra valued parameters $\xi$ are arbitrary functions of the single variable $\th$. Let us finally compute the charge algebra. Let us integrate (\ref{alcs}) at the asymptotic time-fixed boundary $B_\infty$ (remember definition (\ref{loccha})), that is\footnote{The bracket between the two charges is defined as the charge associated to the varied superpotential.}:

\B [Q(\xi),Q(\eta)] := \int_{B_\infty} \d_\eta \U_\xi = Q([\xi,\eta])-\frac{k}{2 \pi} \int_{B_\infty} \ex \eta \cdot \xi \lb{calcs} \E

In terms of the Fourier modes of the charge (since $\th$ is periodic),

\B T_a^p := \int_0^{2 \pi} d \th\ \frac{\d Q(\xi)}{\d \xi^a (\th)}\ \esp^{-ip\th} \lb{ccs} \E

where $p \in \Zed$, this algebra simply becomes a current (affine Kac-Moody) Lie algebra \c{B1} \c{BBGS} with central charge $-k$ (the sign is consistent with our conventions):

\B [T_a^p,T_b^q] = f^c_{ab}\ T_c^{p+q} -ikp\ K_{ab}\ \d^{p+q} \lb{kmal} \E

As usual $f^c_{ab}$ and $K_{ab}$ are the structure constants and the Killing metric of the Lie algebra.

\bigskip

\underline{The diffeomorphism symmetry}:

The Chern-Simons theory is also invariant under an arbitrary Lie derivation parametrized by $\a^\r (x)$:

\B \d_\a \Ab = \li_\a \Ab = (\ex \il_\a +\il_\a \ex) \Ab = \ex \a^\r \il_\r \Ab+ \a^\r \li_\r \Ab \lb{difftran} \E

Here $\il_\r$ and $\li_\r$ are shorthand notations for $\il_{\frac{\p}{\p x^\r}}$ and $\li_{\frac{\p}{\p x^\r}}$ respectively. Now equation (\ref{defu}) becomes:

\B\d \U_\a  = -\frac{k}{2 \pi} \d \Ab \cdot \il_\a \Ab \lb{varucs} \E

The problem is to integrate this equation with our boundary condition $A_t=0$. Unlike in the previous example, this cannot be done in a covariant way and we need to fix the gauge as Ba{\~n}ados \c{B1} showed. The Virasoro algebra was then found by the twisted Sugawara construction (after imposing some additional boundary condition), which gives rive to a classical central extension. 
This central extension in the case of BTZ black hole is nothing but the one computed in \c{BH2}, see \c{B1}.
In a parallel paper \c{JS2}, we rederive this central charge with the Affine gauge theory of relativity \c{JS} without loosing the explicit covariance, since the analogue of (\ref{varucs}) for that case is integrable for the appropriate Dirichlet boundary condition for the metric.

\bigskip

Let us again comment this result:

- Here we can see a drawback of the covariant equation (\ref{defu}), the explicit covariance may have to be broken in order to continue, and for example compute the algebra and the corresponding central charge.

- Borowiec et al. \c{BFF} also computed the superpotential associated to the diffeomorphism invariance by choosing the natural surface term for diffeomorphisms; which is, in analogy to the gravitational case, simply $\il_\a \L_{CS}$. 
The superpotential found (with our convention in the normalisation of the Lagrangian) is just $\U_\a  = -\frac{k}{4 \pi} \Ab \cdot \il_\a \Ab$. It is clear that this superpotential does not satisfy equation (\ref{varucs}). Then in that case the ``natural" choice for the surface term and the ``Hamiltonian inspired" formalism do not coincide.
 The second proposition is dictated by the following facts:

\hspace{.5in} i) Formula (\ref{defu}) is always well defined and does not need choices of surface terms,

\hspace{.5in}  ii) formula (\ref{varucs}) is justified since the
(3-dimensional) Chern-Simons Lagrangian satifies the {\it consistency test}. Moreover
the formula (\ref{varucs}) coincides with the canonical Hamiltonian analysis of Ba{\~n}ados and

\hspace{.5in} iii) if the arbitrary variation of (\ref{varucs}) is precisely another diffeomorphism generated by $\beta$, the variation of the charge $\d_\beta Q(\a) = \int_{B_\infty} \d_\beta \U_\a$ is (on-shell) antisymmetric in $\a$ and $\beta$ (since on-shell $\d_\beta \Ab = \cex \il_\beta \Ab$) as it is expected for an algebraic structure. This does not hold for the superpotential computed by Borowiec et al. \c{BFF}.

\bigskip

\subsection{The higher dimensional Chern-Simons Theory}

It is well known that the usual non-abelian Chern-Simons theory in $3$ dimensions can be generalized to any odd space-time dimension (see for example the detailed analysis of their local degrees of freedom and dynamics in \c{BGH}). 
These theories (and their supersymmetric versions) have been used not only to compute the abelian chiral and non abelian anomalies of higher dimensional currents \c{ZWZ} but also as alternative  (super)gravities (see for example the recent \c{BGM} and \c{TZ} and references therein).

The simplest way to define non-abelian Chern-Simons theories in $(2n+1)$-dimensions is perhaps by its equations of motion:

\B \frac{\d \L_{CS}^{2n+1}}{\d \Ab} =\Fb^n := g_{ab_1 \ldots b_n} \Fb^{b_1}\we \ldots \we   \Fb^{b_n}\  T^a \lb{hcseq} \E

Where $\Fb^n$ is a shorthand notation for $\Fb \we \ldots \we \Fb$ (n times). $\Fb^a$ is, as in the previous example, the curvature of a Lie-valued gauge potential $\Ab^a$ and $g_{ab_1 \ldots b_n}$ an invariant totally symmetric tensor of rank ($n+1$) of the gauge group (that is $g_{ab_1 \ldots b_n}=\mbox{{\it Str}} \left(T_a T_{b_1} \ldots T_{b_n} \right)$, where {\it Str} denotes the totally symmetrized trace and the $T_a$'s form a basis in the adjoint representation of the algebra).

In the following discussion however, we will need some formula for the higher dimensional Chern-Simons Lagrangian. One way to do this is by the nice integral formula \c{ZWZ}:

\B \L_{CS}^{2n+1} := \int^1_0 d\tau \Ab \wed \Fb_\tau^n = g_{ab_1 \ldots b_n} \int^1_0 d\tau \Ab^a \we \Fb^{b_1}_\tau \we \ldots \we \Fb^{b_n}_\tau \lb{hdefl}\E

Where $\Fb_\tau:=\tau \ex\Ab +\tau^2 \Ab \we \Ab$, the integral is just an ordinary integral over the independent variable $\tau$ and the ``dot convention" (see section 4.1, equation (\ref{dot})) was used.

The two local symmetries of these theories are the same as for the previous example:

\bigskip

\underline{The non-abelian gauge symmetry}:

We can compute in a straightforward way the variation of the
superpotential (equations (\ref{defv}) and (\ref{defu})) associated to this symmetry by use of (\ref{ymtran}) and (\ref{hcseq}):

\B \d \U_\xi = - n\ \xi \cdot \d \Ab \we \Fb^{n-1} =- n \ \xi^a g_{abc_2 \ldots c_n} \d \Ab^b \we \Fb^{c_2}\we \ldots \we   \Fb^{c_n} \lb{varhcs} \E

First note that the above equation can be obtained from the  
Hamiltonian analysis (following the Regge-Teitelboim prescription) for
conserved charges\footnote{See
for instance equation (7.3) of the second reference of \c{BGH} for the
special 5-dimensional case.}. 
The {\it consistency test} given in the appendix has to be verified for
these higher dimensional Chern-Simons theories since the Lagrangian
(\ref{hdefl}) is not of the simple type (\ref{Lag1or}). A
straightforward computation of $\delta \f  \frac{\delta
W^{\mu }_{\xi}}{\delta \f}$ (using either its definition (\ref{corect}) or
the simpler formula (\ref{simcheck})) gives:

\begin{equation}\label{conchhcs}
n \delta \Ab \wed \delta_{\xi} \Ab \we \Fb ^{n-1}
\end{equation}

The vanishing of this expression at infinity will be checked after the
following discussion on boundary conditions (see the arguments after equations (\ref{hbc}-\ref{hbcp})).

Now, as in the simplest 3-dimensional Chern-Simons theory, the equation (\ref{varhcs}) can be integrated in a covariant way. To do that, we will need to choose the boundary conditions for this theory. First, equation (\ref{varhcs}) looks very much like the variation of a ($2n-1$)-dimensional Chern-Simons Lagrangian (\ref{hcseq}). In view of (\ref{hdefl}), we are thus naturally tempted to try:

\B \U_\xi =-n\  \xi \cdot \int^1_0 d\tau \Ab \we \Fb^{n-1}_\tau := -   \L^{2n-1}_{CS} (\xi) \lb{hcsu} \E

However, due to the fact that the gauge parameter $\xi$ depends {\it a priori} on $x$ (and so does the ``effective" (rank $n$)-symmetric tensor $g(\xi)_{bc_2 \ldots c_n} :=n \ \xi ^a g_{abc_2 \ldots c_n}$ of the ($2n-1$)-dimensional Chern-Simons Lagrangian), the variation of (\ref{hcsu}) is, up to an exact $(2n-1)$-form which is physically irrelevant (see discussion after equation (\ref{startpoint})), precisely (\ref{varhcs}) plus the non trivially vanishing term:

\B - n\ex \xi\wed \d \Ab \we \frac{\p \int^1_0  d\tau \left( \Ab \wed \Fb_\tau^{n-1}\right)}{\p \ex \Ab}  = -n(n-1)\ex \xi\wed  \d \Ab \we \int^1_0  d\tau \tau\ \Ab \we \Fb_\tau^{n-2}\lb{ext} \E

The point now is to argue that this term will vanish after making use of the variational boundary conditions. The variational principle on the Chern-Simons Lagrangian (\ref{hdefl}) requires that the boundary term

\B \int_{\Sigma_\infty}  \d \Ab \wed \frac{\p \L_{CS}^{2n+1}}{\p \ex \Ab} = n\int_{\Sigma_\infty} \d \Ab \wed \int^1_0  d\tau\tau\  \Ab \we \Fb_\tau^{n-1} \lb{bthcs} \E

\noindent vanishes. A way to satisfy this condition is quite similar to the 3-dimensional case (\ref{cscond}) . Let us pursue the analogy and use some ``cylindrical type" coordinate system $(t,r,\th_1,\ldots,\th_n,\phi_1,\ldots ,\phi_{n-1})$. 
As before, $t$ and $r$ parametrize the time direction and the normal to the asymptotic boundary $\Sigma_\infty$, now taken isomorphic to $\Re\times S^{2n-1}$.
Now, the $\th$'s together with the $\phi$'s  parametrize the asymptotic sphere $S^{2n-1}$. A way to do that which will be useful for the following is by making use of $n$ periodic variables, say the $n$ $\th$'s, $\th \in [0,2 \pi]$. 
This parametrization is always possible due to the following observation: the $S^{2n-1}$ sphere can be embedded in $R^{2n}$ and therefore can be parametrized by $n$ pairs of polar variables (i.e. $n$ planes): $(\th_i,r_i)$, $i=1 \ldots n$, subject to the restriction $\sum_{i=1}^{n} r_i^2 =C^{t}$. Thus our $n$ $\th$'s are nothing but these $n$ polar angles.

Now, a natural generalization of the $A_t=0$ boundary condition is the vanishing of $A_t$ and of the $A_\phi$'s. In other words, the only non vanishing $\Ab$'s at the boundary will be the $A_\th$'s. It is easy to convince oneself that this will imply the vanishing of (\ref{bthcs}). So we may impose:

\B A_t= A_{\phi_1} = \ldots = A_{\phi_{n-1}} =0\hspace{.5in}  \mbox{at the boundary} \lb{hbc} \E

As usual this implies some asymptotic constraints on the gauge parameter (see again (\ref{bcres})) as in the 3-dimensional case (\ref{cscon}):

\B \p_t \xi = \p_{\phi_1} \xi= \ldots = \p_{\phi_{n-1}} \xi =0\hspace{.5in} \mbox{at the boundary} \lb{hbcp} \E

And so the gauge parameter cannot depend on time or $\phi$'s coordinates. These $\phi$'s coordinates (or their respective vector fields, the $\frac{\p}{\p \phi}$'s) will be called the {\it vanishing} coordinates (or {\it vanishing} directions), in contrast with the {\it periodic} $\th$'s coordinates (directions). 
Now the vanishing of the integral on the asymptotic sphere $S^{2n-1}$ of (\ref{ext}) follows trivially from (\ref{hbc}) and (\ref{hbcp}). 
Equation (\ref{ext}) is a wedge product of $\d \Ab$ and at least
$(n-1)$ $\Ab$'s (at least one per $\Fb$) and one $\ex \xi$. Since the
asymptotic manifold is ($2n-1$)-dimensional, at least one of the
$\Ab$'s or $\ex \xi$ will be in one of the {\it vanishing} directions,
which will imply the vanishing of (\ref{ext}). Now, the vanishing of
equation (\ref{conchhcs}) at infinity (the {\it consistency test}) can be verified using
similar arguments. Thus, this fully justifies the use of (\ref{varhcs}) for
computing the superpotential.

Equation (\ref{hcsu}) gives a really nice picture as of the superpotential of a $(2n+1)$-Chern-Simons Lagrangian as a $(2n-1)$-Chern-Simons Lagrangian. 
We finally would like to point out that the integrated superpotential (\ref{hcsu}) coincides with the one that could be computed with the cascade equations techniques if we were to choose the following boundary\footnote{The variation of the ($2n+1$)-dimensional Chern-Simons Lagrangian under an non-abelian gauge symmetry has been carried out in \c{ZWZ}.} term:

\B \d_\xi \L^{2n+1}_{CS} = \ex \S_\xi \hspace{.2in} \mbox{with} \hspace{.2in} \S_\xi= n \int_0^1 d\tau (\tau-1) \ex \xi \wed \Ab \we \Fb_\tau^{n-1} \lb{hsute} \E

Of course, another choice of surface term would give another superpotential as was emphasized in section 2.

\bigskip

Let us return to the superpotential (charge) algebra. To find it, we can first observe that due to the boundary conditions (\ref{hbc}) and (\ref{hbcp}), the physical charge (\ref{loccha}) (with (\ref{hcsu})) simplifies to:

\B Q(\xi) = \int_{B_\infty} \U_\xi = -n \int_{B_\infty} \xi \cdot \int_0^1 d\tau \Ab\we (\tau \ex \Ab)^{n-1}\lb{defusi} \E

A simple way to understand this is the following: All the other terms derived from (\ref{hcsu}) will contain more than $n \ \Ab$'s. 
In that case, at least one of these $\Ab$'s will be in one of the {\it vanishing} directions (since this expression is integrated on $S^{2n-1}$, also called $B_\infty$), which will imply by (\ref{hbc}) the vanishing of this term. Thus, the only {\it possibly} non-vanishing term is (\ref{defusi}), which contains exactly $n\ \Ab$'s (the ones in the {\it periodic} directions).

We can rewrite (\ref{defusi}) in an even more compact form after making the allowed replacements $\ex \Ab \rightarrow \Fb$ (for the same reasons as before) and integrating over $\tau$:

\B Q(\xi) =- \int_{B_\infty} \xi \cdot \Ab\we\Fb^{n-1} \lb{hcc} \E

We are now ready to proceed as in previous examples: Let us vary equation (\ref{hcc}) with respect to a gauge symmetry parametrized by $\eta$, always keeping $\xi$ fixed (and using the same trick as, for example (\ref{ctrick})):

\B \d_\eta Q(\xi) = -\int_{B_\infty}\left[ \d_\eta \left(\xi\cdot  \Ab\we\Fb^{n-1} \right) - \d_\eta \left(\xi\right) \cdot \Ab\we\Fb^{n-1} \right] \lb{calcc} \E

The last term in the rhs is just $Q([\xi,\eta])$. The first term
has no floating gauge index and so would have been a gauge scalar if
all the fields had transformed homogeneously under the gauge
group. This is the case for the curvature but not for the gauge
potential, which gives an extra term $\ex \eta^a$ as in section
4.2. Then, (\ref{calcc}) can be rewritten as:

\B[Q(\xi),Q(\eta)] :=\d_\eta Q(\xi) =  Q([\xi,\eta]) - \int_{B_\infty} \xi \cdot \ex \eta \we \Fb^{n-1} \lb{alg}\E

Now, in the case where $n\geq 2$, the second term of the rhs of
(\ref{alg}) vanishes\footnote{Recall that for $n=1$, this term was just the
central charge contribution to the algebra (\ref{calcs}).}.
This can be checked by replacing $\Fb$ by $\ex \Ab$ (see previous
discussion after equation (\ref{defusi})) and integrating by parts,
together with the boundary conditions (\ref{hbc}) and (\ref{hbcp}).
This algebra is another example with respect to what was found in \c{LMNS} and
\c{BGH} and references therein (the so-called $\mbox{WZW}_{2n}$
algebra). In these papers, the special assumption that the gauge group
had a $U(1)$ factor was used. Moreover different boundary conditions
(others than (\ref{hbc})) were imposed, which allowed the presence of
a central charge in the algebra.

It is interesting to note the similarity which exists between our result and  the
case of general relativity with a negative cosmological constant \c{BH2}: The
asymptotic symmetry algebra is equipped with a central charge only in
$3$ spacetime dimensions and no more.

To complete the analysis, we can as in the $3$-dimensional case expand this algebra in terms of $n$-dimensional Fourier modes: From (\ref{hbcp}), the parameters $\xi$ will depend only on the $n$ $\th$'s variables (the {\it periodic} variables). Using the simple vector notation $\vec{\th} := (\th_1,\ldots,\th_n)$, this becomes $\xi=\xi(\vec{\th})$, that is, we can generate charges with any Lie algebra valued function of $n$ {\it periodic} variables. Let us then define from (\ref{hcc}) the $n$-dimensional Fourier modes:

\B T^{\vec{p}}_a := \int_{0}^{2 \pi} d\vec{\th}^n\ \frac{\d Q(\xi)}{\d \xi^a (\vec{\th})}\ \esp^{-i\vec{p} .  \vec{\th}} \lb{fcoe} \E

Now, $\vec{p}$ is an $n$-dimensional vector of the integer lattice
$\Zed^n$ and ``$.$" symbolizes the $n$-dimensional Euclidean scalar
product. Then the algebra (\ref{alg}) reduces to (for $n\geq 2$)

\B [T^{\vec{p}}_a,T^{\vec{q}}_b] = f^c_{ab} T^{\vec{p}+\vec{q}}_c  \lb{algn}\E

This algebra verifies the Jacobi identity and generalizes in a natural way the current algebra (on $S^1$) of the well known 3-dimensional case (\ref{kmal}).

\bigskip

\underline{The diffeomorphism symmetry}:

The formula for the variation of the superpotential is completely analogous to the 3-dimensional case:

\B \d \U_\a = - n \d \Ab^a g_{abc_2 \cdots c_n} \il_\a \Ab^b \we \Fb^{c_2}\we \cdots \we   \Fb^{c_n} = -n \il_\a \Ab\cdot \d \Ab \we \Fb^{n-1} \lb{varhcs2} \E

The problem is also the same. This equation cannot be integrated in a covariant way and only a careful analysis of boundary and gauge fixing conditions could allow us to obtain some integrated superpotential.

Finally, it can also be checked explicitly from (\ref{varhcs2}) that if the variation is another  diffeomorphism (\ref{difftran}), the associated charge $\d_\beta Q (\a)$ will be (at least on-shell) antisymmetric in $\beta$ and $\a$ as expected. This again would not be the case if the superpotential were computed with the cascade equations formalism with the ususal choice for the surface term, that is $\S_\a =\il_\a \L_{CS}^{2n+1}$.

\bigskip

\subsection{The Abelian p-forms Chern-Simons Theory}

The abelian p-form Chern-Simons theory was treated in great detail in \c{BHIV}. This is yet another example of topological (in the sense that no metric is needed) field theory, just as the previous two examples. Let $\Bb^a$ ($a=1 \ldots N$) a family of N abelian $p$-form gauge fields, with corresponding curvatures $\Gb^a := \ex \Bb^a$. The Lagrangian is \c{BHIV}:

\B \L_p^n := \frac{1}{n+1} g_{ab_1 \ldots b_n} \Bb^{a} \we \Gb^{b_1} \we \ldots \we \Gb^{b_n} = \frac{1}{n+1} \Bb\wed \Gb^n \lb{plag} \E

\noindent $n$ is the ``order" of the $p$-form Chern-Simons theory and depends on $p$ and $D$, the space-time dimension, through the simple equation $D=p+n (p+1)$. The rank-$(n+1)$ invariant tensor $g_{ab_1 \ldots b_n}$ will be assumed to be totally (anti)symmetric if $p$ is (even)odd. The ``dot convention" is completely analogous to the previous examples: 
it simply means ``each time a dot appears, contract all the $^a$'s indices with the invariant totally (anti)symmetric tensor $ g_{ab_1 \ldots b_n}$".

The equations of motion corresponding to (\ref{plag}) are the familiar ones:

\B \frac{\d \L_p^n}{\d \Bb^a} = g_{ab_1 \ldots b_n} \Gb^{b_1} \we \ldots \we \Gb^{b_n} \lb{eqmop} \E

Again, we will search for the superpotential of the two local symmetries of this theory:

\bigskip

\underline{The p-form abelian gauge symmetry}:

The superpotential associated to the gauge symmetry $\d_{\xib} \Bb^a = \ex \xib^a$ (now $\xib^a$ are $N$ abelian $(p-1)$-forms) will be given by:

\B \d \U_\xi =-n\ \d \Bb^b \we \xib^a  g_{abc_2 \ldots c_n} \Gb^{c_2} \we \ldots \we \Gb^{c_n} =-n\ \d \Bb \wed \xib \we \Gb^{n-1} \lb{pvaru} \E

As in the previous example, the Lagrangian (\ref{plag}) is not in
general of the type (\ref{Lag1or}). That means that the {\it
consistency test} of the appendix has to be verified by hand. Using either
(\ref{corect}) or (\ref{simcheck}), we find that
\begin{equation}\label{c22}
n \delta \Bb \wed \delta _{\xi}\Bb \we \Gb^{n-1}
\end{equation}
has to vanish at infinity. This will be checked in the following,
after equation (\ref{pcsb2}).

The integration of the equation (\ref{pvaru}) can be done after some careful analysis of the variational principle boundary conditions. The natural guess,

\B \U_\xi = -\Bb\wed \xib \we \Gb^{n-1} \lb{psp} \E

\noindent suffers from the same problem, namely the variation of (\ref{psp}) gives precisely (\ref{pvaru}) plus the {\it a priori} non vanishing term:

\B -(n-1)\d \Bb \wed \ex \xib \we \Bb \we \Gb^{n-2} \lb{pplus} \E

Let us now check that this term vanishes if we solve the boundary conditions dictated by (\ref{plag}), that is:

\B \frac{n}{n+1} \int_{\Sigma_\infty} \d \Bb \wed \Bb \we \Gb^{n-1} = 0 \lb{pbc} \E

To solve this boundary constraint we use again the cylindrical-type coordinates $(t,r,\th_1,\ldots,\th_{D-n-1},\phi_1,\ldots,\phi_{n-1})$, with $\frac{\p}{\p t}$ pointing in the time direction and $\frac{\p}{\p r}$ being normal to the boundary $\Sigma_\infty$. In complete analogy with the higher dimensional Chern-Simons theories, we can choose the following boundary conditions:

\B \il_t \Bb = \il_{\phi_1} \Bb= \ldots = \il_{\phi_{n-1}} \Bb= 0 \hspace{.3in} \mbox{at the boundary} \lb{pcsb} \E

Here $\il_t$ is a shorthand notation for $\il_{\frac{\p}{\p t}}$ and so on. As in section 4.3 we will call the $(n-1)$ $\phi$'s coordinates (or their corresponding vector fields) the {\it vanishing} coordinates (directions). The conditions (\ref{pcsb}) impose on the $\xib$ parameters the following conditions:

\B \il_t \ex \xib= \il_{\phi_1} \ex \xib= \ldots = \il_{\phi_{n-1}} \ex \xib= 0 \hspace{.3in} \mbox{at the boundary} \lb{pcsb2} \E

Now again (\ref{pcsb}) and (\ref{pcsb2}) imply the vanishing of (\ref{pplus}). On the asymptotic $(D-2)$ sphere (i.e. $B_\infty$, parametrized by the $\th$'s and the $\phi$'s), there exist $(n-1)$ vanishing directions. At worst, $(n-2)$ of these directions will correspond to the $(n-2)$ (exterior) derivatives of the $\Gb^{n-2}$ term in (\ref{pplus}). 
The last one will correspond either to a $\Bb$ or to the $\ex \xib$ term, and so the whole expression will vanish. 
Similar arguments can be used to prove the vanishing of (\ref{c22}),
and so the correctness of (\ref{pvaru}).

We obtain again the fact that the superpotential of a $(p,n)$-Chern-Simons theory (\ref{psp}) is given by a $(p,n-1)$-Chern-Simons Lagrangian formula (\ref{plag}), with an ``effective" rank-$n$ $(p-1)$-form totally (anti)symmetric tensor $g(\xib)_{bc_2\ldots c_n} = \xib^a\ g_{abc_2\ldots c_n}$. 
Note again that the superpotential (\ref{psp}) can be computed with the cascade equations techniques, with as specific choice for the boundary term the analogue of (\ref{hsute}), that is $\S_{\xib} = \frac{(-1)^p}{n+1} \ex \xib \wed \Bb \we \Gb^{n-1}$.

Finally, the charge algebra can be computed exactly in the same way as in previous examples. We only give the final result:

\B[Q(\xib),Q(\etab)] :=\d_{\etab} Q(\xib) =  - \int_{B_\infty} \xib \cdot \ex \etab \we \Gb^{n-1} \lb{palg}\E

In analogy with the $p=1$ non-abelian Chern-Simons theories, a central
charge appears in the abelian algebra (\ref{palg}) only for $n=1$. In fact,
for $n\geq 2$, the $\Gb^{a}=\ex \Bb^{a}$ can be integrated by
part in the rhs of (\ref{palg}). The resulting
expression then vanishes due to the boundary conditions (\ref{pcsb})
and (\ref{pcsb2}).

\bigskip

\underline{The diffeomorphism Symmetry}:

The variation of the superpotential associated to the diffeomorphism symmetry is given by:

\B \d \U_\a = -k \d \Bb \wed \il_\a \Bb \we \Gb^{k-1} \E

Again the integration of this equation cannot be done without fixing the gauge.

\bigskip

\section{Conclusions and Perspectives}

The principal result of this paper is equation (\ref{defu}) which gives a new covariant way to compute the superpotential, and then the charges associated to gauge symmetries. 
Moreover no canonical split of space and time is needed. This formula
is also intimately related to the Noether techniques (or cascade
equations) as expected for a Lorentz invariant formula. Finally, a
{\it consistency test} is given in the appendix to justify such
a proposal.

This new definition of superpotential has been tested on the
Yang-Mills and Chern-Simons type theories (the examples of gravity in
the {\it affine} gauge formulation \c{JS} and supergravity will be studied
in two forecoming paper \c{JS2} and \c{HJS} respectively). It is in perfect agreement with the Hamiltonian and other results in $3$-dimensional Chern-Simons theories and allows to compute the higher dimensional generalizations. 
One of the drawbacks is that a superpotential for the diffeomorphism symmetry of topological Chern-Simons theories can not be written in a covariant way. In that case, we cannot escape from the gauge fixing analysis which breaks the explicit Lorentz invariance. We finally guess that a supersymmetric extension of the above results should be possible.

The charge algebras (with their eventual central extensions) were also defined in a covariant way by the variation of the superpotential under another gauge symmetry. This procedure is inspired from the canonical formalism. In that case, a nice theorem \c{BH1} (see also \c{Ar})
guarantees the homomorphy (up to a central charge) between the Lie-gauge algebra and the associated charge algebra. It would be interesting to know if this theorem can be generalized to the covariant case. The examples treated in section 4 seem to point in that direction.

\bigskip

{\bf Acknowledgments.}
\bigskip

I would like to thank B. Julia for his constant encouragements and critical but constructive discussions. I am most grateful to M. Henneaux for answering my numerous questions and for motivating  me to look at Chern-Simons theories.  It is also a pleasure to thank E. Cremmer for stimulating conversations.

I am also grateful to the EU TMR contract FMRX-CT96-0012 for travel support.

\section*{Appendix: A consistency test}
\renewcommand{\theequation}{A.\arabic{equation}}
\setcounter{equation}{0}

The purpose of this appendix is to derive a {\it consistency test} for
our proposal (\ref{defu}). We will also give a rigourous proof that
this {\it consistency} always holds for a theory in a first
order formulation (in a sense to be precised).

\subsection*{A.1 The Test}

The idea goes as follow: since $J^{\mu}_{\xi}$ vanishes at infinity
(recall equation (\ref{vanj})), so does an arbitrary variation of it. Thus,
at infinity, equation (\ref{startpoint2}) has to vanish, that
is\footnote{Note that we are now writing explicitly field indices with uppercase latin letters.}:

\B \d \f^{A} \ \frac{\d W_\xi^\m}{\d \f^{A}}  + \p_\n
V^{\m\n}(W_\xi^\m, \d \f )+ \p_\n \d U^{\m\n}_\xi =0 \hspace{.2in} \mbox{at infinity} \label{stdem} \E

A {\it consistency test} for (\ref{defu}) is thus to verify that the first term in equation
(\ref{stdem}) vanishes due to the boundary conditions needed to
have a well defined variational principle for the Lagrangian. That is 

\begin{equation}\label{corect}
\d \f^{A} \ \frac{\d W_\xi^\m}{\d \f^{A}} = 0 \hspace{.2in}\mbox{at infinity}
\end{equation}
\noindent with the definition (\ref{defw})
\begin{equation}\label{rdefw}
 W_\xi^\m :=\Delta^{\mu A}_{a} \xi^{a} \frac{\delta L}{\delta \f^{A}}
\end{equation}
In fact, if (\ref{vanj}) and (\ref{corect}) are satisfied for some
theory (with chosen boundary conditions), then (\ref{defu}) holds.

\bigskip 

In practice, it is not always so simple to evaluate (\ref{corect}) for
a given theory; that is why we will give an equivalent expression for
it, that is much simpler to compute. In addition, we give an explicit  proof of
the antisymmetry in $\mu$ and $\nu $ of $V_{\xi}^{\mu\nu}$ defined in
(\ref{defv}). 

The Lagrangian $L$ is invariant under a
gauge symmetry, parametrized by an infinitesimal gauge parameter
$\xi^{a}$, with the transformation laws of the fields given by equation
(\ref{tran1}). The associated equations of motion $E_{A}:=\frac{\delta
L}{\delta \f^{A}}$ satisfy some
Noether identities due to the gauge invariance (see comment after equations
(\ref{idcas1}-\ref{idcas3})), namely:

\B \delta_{\xi} \f^{A} E_{A} = \partial_{\mu} \left(
\xi^{a}\Delta_{a}^{\mu A}
E_{A} \right) \lb{noidfo}\E

The fact that the rhs of (\ref{noidfo}) is a total derivative
implies that its Euler-Lagrange variation vanishes:

\B  \frac{\delta }{\delta \f^{B}} \left( \delta_{\xi} \f^{A} 
E_{A}\right) = 0 \label{cons1} \E

Now, as in section 2, we will use the abelian restriction \c{JS} given
by equation (\ref{abelres}) to
extract information hidden in equation (\ref{cons1}). In the
spirit of cascade equations \c{JS}, we look for the term proportional to
$\e$, $\partial_{\mu} \e$ and $\partial_{\mu} \partial_{\nu}\e$ in (\ref{cons1}). So using the transformation law (\ref{tran1}) we
find after some little algebra that:

\begin{equation}
\e\  \left[ \frac{\delta }{\delta \f^{B}}\left( \delta_{\xi_{0}} \f^{A}
E_{A}\right) \right] = 0\label{cas111}
\end{equation}
\begin{equation}
\partial_{\mu} \e \  \left[ \frac{\delta }{\delta \f^{B}}\left(\Delta^{\mu
 A}_{a} \xi^{a}_{0} E_{A} \right)-
\frac{\partial}{\partial\partial_{\mu}\f^{B}}\left(\delta_{\xi_{0}}\f^{A}
E_{A}\right)\right] = 0 \label{eqcasc}
\end{equation}
\begin{equation}
\partial_{\mu}\partial_{\nu} \e \ \left[ -\frac{\partial
}{\partial\partial_{\nu}\f^{B}}\left( \Delta^{\mu
 A}_{a} \xi^{a}_{0} E_{A}\right) \right] =0 \label{antisym}
\end{equation}
\noindent where the subscript $_{0}$ is defined by (\ref{abelres}) and
will be
omitted in the following for simplicity. We used the
restriction that the equations of motion and $\delta_{\xi}\f^{A}$
depend at most on first derivatives of the fields.

Since $\e$ is arbitrary, the expressions (\ref{cas111}-\ref{antisym}) have to vanish
identically. Let us discuss each equation separately: 
\begin{enumerate}
\item The first equation (\ref{cas111}) is nothing but equation
(\ref{cons1}) with an additional $_{0}$ subscript.

\item When contracted with $\delta \f ^{B}$, the first term of
equation (\ref{eqcasc}) is
nothing but the lhs of (\ref{corect}) (remember the definition
(\ref{rdefw})). Then, we find that the {\it consistency test}
(\ref{corect}) simplifies to:

\begin{equation}\label{simcheck0}
\delta \f^{B} \frac{\partial ( \delta_{\xi}\f^{A})}{\partial\partial_{\mu}\f^{B}} E_{A} +
\delta \f^{B} \delta_{\xi}\f^{A}  \frac{\partial
E_{A}}{\partial\partial_{\mu}\f^{B}} = 0 \hspace{.2in} \mbox{at infinity}
\end{equation}

The first term in the above equation vanishes on-shell and so can be
eliminated classically. Therefore, the {\it consistency test}
(\ref{corect}) can be rewritten on-shell as: 
\begin{equation}\label{simcheck}
\delta \f^{B} \delta_{\xi}\f^{A}  \frac{\partial
E_{A}}{\partial\partial_{\mu}\f^{B}} = 0 \hspace{.2in} \mbox{at infinity}
\end{equation}

\item When contracted with $\delta \f ^{B}$, The last equation (\ref{antisym}) is nothing but $-V_{\xi}^{\mu
\nu }$ defined by (\ref{defv}), contrated with the symmetric tensor
(in $\mu $ and $\nu $) $\partial_{\mu }\partial_{\nu } \e$. Then, the
vanishing of equation (\ref{antisym}) implies the antisymmetry (in
$\mu $ and $\nu $) of the superpotential defined by (\ref{defu}).  

\end{enumerate}

\subsection*{A.2 A proof for the main proposal} 

We will end this appendix by giving an explicit proof for equation
(\ref{corect}) in the simple case where the Lagrangian is of first
order type, in the sense that:

\begin{equation}\label{Lag1or}
L = f^{\mu}_{A} \partial_{\mu}\f^{A}   + g
\end{equation}
Here $f^{\mu}_{A}=f^{\mu}_{A} (\f)$ and $g=g(\f)$ are functional only of
the fields $\f^{A}$ but not of their derivative. This restriction is
sufficiently general to allow Yang-Mills theories and
(super)gravities in their first order formulation. Chern-Simons theories
are more involved and cannot be written for $D\geq 5$ in the simple
form (\ref{Lag1or}). However, it is a straightforward computation to
verify that these theories satisfy (\ref{corect}) for the boundary
conditions given in section 4.

The boundary conditions are defined by the
variational principle for the Lagrangian (\ref{Lag1or}), that is:

\B \int_{\Sigma_{\infty}} \delta \f^{A} f^{\mu}_{A} dS_{\mu}\equiv  0
\label{boudco11}\E 

\noindent where $\Sigma_{\infty}$ denotes again the boundary (at infinity) of the spacetime
manifold. Moreover, if we add a total divergence to the
Lagrangian (\ref{Lag1or}), the boundary problem to be satisfied is
shifted by:
\begin{equation}\label{shift}
L\rightarrow L+\partial_{\mu} k^{\mu} (\f) \hspace{.2in} \mbox{then}
\hspace{.2in} f^{\mu}_{A} \rightarrow f^{\mu}_{A} + k^{\mu}_{,A}
\end{equation}
\noindent where the $_{,A}$ denote a partial derivative with respect to the
field $\f^{A}$.

The equations of motion for the Lagrangian (\ref{Lag1or}) are:

\B E_{A}:=\frac{\delta L}{\delta \f^{A}}= \left(
f^{\mu}_{B,A}-f^{\mu}_{A,B}\right) \partial_{\mu} \f^{B} + g_{,A}
\label{eqmo22} \E

Using now the above equations of motion we obtain that on-shell the
{\it consistency test} given by equation
(\ref{simcheck}) becomes:

\begin{equation}\label{proof}
\delta_{\xi}\f^{A} \delta \f^{B} \left(
f^{\mu}_{B,A}-f^{\mu}_{A,B}\right) =\delta \f^{A} \delta_{\xi}
f^{\mu}_A  -  \delta_{\xi}\f^{A}  \delta f^{\mu}_{A} 
\end{equation}

Now this expression clearly vanishes at infinity due to the boundary
conditions (\ref{boudco11}). Note also that the addition of a total
divergence to the Lagrangian (see equation (\ref{shift})) does not
modify expression (\ref{proof}).

As a final comment, let us point out some relation with the covariant
Hamiltonian formalism of Witten-Crnkovic-Zuckerman
\cite{WCZ}: Equation (\ref{boudco11}) is the so-called Noether form 
integrated on $\Sigma_{\infty}$. The exterior ``differential'' $\delta
$ (in the
space of classical solutions to field equations) of this
equation is the so-called ``symplectic current'' or
Witten-Crnkovic-Zuckerman 2-form, but now integrated on
$\Sigma_{\infty}$. What we have shown is that (\ref{corect}) is
nothing but the  Witten-Crnkovic-Zuckerman 2-form contracted along a
symmetry direction $\delta _{\xi} \f^{A}$, which vanishes when  integrated on $\Sigma_{\infty}$.

\end{document}